\documentclass{mn2e}
\usepackage{graphicx}

\def\lesssim{\buildrel < \over {_{\sim}}}
\def\gtrsim{\buildrel > \over {_{\sim}}}
\def\HH{H$_2$ }
\def\msun{{\rm M_\odot}}

\def\cm{{\rm cm}}
\def\percmc{{\rm cm^{-3}}}
\def\gram{{\rm g}}
\def\kelvin{{\rm K}}
\def\sigmav{\langle\sigma v\rangle}

\begin{document}

\title[First star formation with DM annihilation]{First star formation with dark matter annihilation}

\author[]{E. Ripamonti$^{1,2}$, F. Iocco$^{3}$, A. Ferrara$^{4}$, 
R. Schneider$^{5}$, A. Bressan$^{6,7,8}$, P. Marigo$^{9}$\\
$^{1}$Universit\`a degli Studi di Milano-Bicocca, Dip. di Fisica ``G. Occhialini''; Piazza della Scienza 3, I-20126 Milano, Italy\\
$^{2}$Universit\`a degli Studi dell'Insubria, Dip. di Scienze Chimiche, Fisiche e Naturali; Via Valleggio 12, Como, Italy\\
$^{3}$Institut d`Astrophysique de Paris, UMR 7095-CNRS Paris, Universit\'e Pierre et Marie Curie, 98bis Bd Arago, F-75014, Paris, France\\
$^{4}$Scuola Normale Superiore, Piazza dei Cavalieri 7, Pisa, Italy\\
$^{5}$INAF/Osservatorio Astrofisico di Arcetri, Largo Enrico Fermi 5, Firenze, Italy\\
$^{6}$SISSA; Via Beirut 4, Trieste, Italy\\
$^{7}$INAF/Osservatorio Astronomico di Padova; Vicolo dell'Osservatorio 5, Padova, Italy\\
$^{8}$INAOE; Luis Enrique Erro 1, 72840, Tonantzintla, Puebla, Mexico\\
$^{9}$Universit\`a degli Studi di Padova, Dip. di Astronomia; Vicolo dell'Osservatorio 2, I-35122 Padova, Italy}
\pagerange{\pageref{firstpage}--\pageref{lastpage}} \pubyear{2009}
\maketitle

%\label{firstpage}

\begin{abstract}
We include an energy term based on Dark Matter (DM)
self-annihilation during the cooling and subsequent collapse of the metal-free gas, 
in halos hosting the formation of the first stars in the Universe.  We have
found that the feedback induced on the chemistry of the cloud {\it
does} modify the properties of the gas throughout the collapse.
However, the modifications are not dramatic, and the typical Jeans
mass within the halo is conserved throughout the collapse, for all the
DM parameters we have considered.  This result implies that the
presence of Dark Matter annihilations does not substantially modify the Initial
Mass Function of the First Stars, with respect to the standard case in
which such additional energy term is not taken into account.  We have
also found that when the rate of energy produced by the DM annihilations and
absorbed by the gas equals the chemical cooling (at densities yet far
from the actual formation of a proto--stellar core) the structure does
{\it not} halt its collapse, although that proceeds more slowly by a
factor smaller than few per cent of the total collapse time.
\end{abstract}


\section{Introduction}
In the currently favoured $\Lambda$CDM cosmological model, the bulk of the matter component is believed to be made of (so far)
electromagnetically undetected particles, commonly dubbed Dark Matter (DM). 
Although the evidence for the existence of DM is compelling
on different scales, yet its nature is unknown, and many particle models beyond 
the standard one have been proposed in the literature
as DM candidates. We address the reader to a recent review of observational 
evidence and particle candidates for DM (e.g. Bertone, Hooper \&
Silk 2005), and will concentrate in this paper on a particular class of candidates, 
i.e. Weakly Interacting Massive Particles (WIMPs). 
Many WIMP DM models are stable (under the conservation
of the suitable symmetry, for each model) and hence
do never decay into standard model particles.
However, in many of these very same models the
WIMPs are Majorana particles, thus carrying the
remarkable property of  being self-annihilating;
the value of the self--annihilation cross section, 
arising naturally in many WIMP models, reproduces the 
dark matter relic abundance required by the $\Lambda$CDM cosmology, 
if the mass scale of WIMPs is within the GeV/TeV scale and they are to be
thermally produced in the early Universe. We adopt this as 
a benchmark scenario for our paper, and will often refer to it as a ``Vanilla WIMP''.

The actual DM distribution in the local Universe is such that even 
in the densest regions (e.g. galactic nuclei and black hole surroundings) 
from which the annihilation signal could be in principle detected, the energy released by WIMP DM annihilations (hereafter, DMAs) is 
only a negligible fraction of the one associated with standard gas processes.
%\footnote{Although gamma-rays and secondary cosmic rays  could be detected by current or planned experiments such as FERMI or AMS--2.}. 
This implies that, locally, DM affects the host system 
almost uniquely through its gravitational effects, perhaps with the only possible exception of peculiar locations, such as the central 
parsec of the Milky Way (Fairbairn, Scott \& Edsjo 2008,  Scott, Edsjo \& Fairbairn 2009, Casanellas \& Lopes 2009). 
The effects of annihilating (or decaying, a scenario we do not consider here) 
DM upon the evolution of the intergalactic medium (IGM) at high redshift have 
been thoroughly studied (e.g. Chen \& Kamionkowski 2004; Padmanabhan \& Finkbeiner 2005; Mapelli \& Ferrara 2005; Mapelli, Ferrara \& Pierpaoli 2006; Furlanetto, Oh \& Pierpaoli 2006; Zhang et al. 2006; Ripamonti, Mapelli \& Ferrara 2007a; Vald\'es et al. 2007; Shchekinov \& Vasiliev 2007),
and are now believed to be small, except perhaps in the case of an extremely high clumping factor (Myers \& Nusser 2008; Chuzhoy 2008; Natarajan \& Schwarz 2008, Lattanzi \& Silk 2009), if one takes into consideration a standard, Vanilla WIMP scenario.  

The effects of DMAs upon primordial star-formation might be more significant. 
As the IGM could be heated by the energy deposition from DMAs, 
its temperature might in principle exceed the virial temperature of the 
smallest halos with the result of quenching gas accretion onto them. 
This effect has been shown to be unimportant  by 
Ripamonti, Mapelli \& Ferrara (2007b, hereafter RMF07). 
However, DMAs are expected to become 
more important as the collapse proceeds to protostellar scales (Ascasibar 2007). 
Spolyar, Freese \& Gondolo (2008; hereafter SFG08) found that during the 
protostellar collapse of the first (Pop III) stars, the energy released by DMAs 
and absorbed by the gas could compensate (or even overcome) the 
radiative cooling of the gas.  The increasing importance of such process 
arises from the combined enhancement during the collapse of DM density 
(due to gravitational dragging) and gas optical depth, implying a higher 
annihilation luminosity and absorption by the gas. The final phases of the 
collapse, after the formation of a hydrostatic core for gas central densities 
$n_{\rm c} \equiv \rho_{\rm c}/m_{\rm p} > 10^{18}\,\percmc$ (where $\rho_{\rm c}$ 
is the central baryonic density, and $m_{\rm p}$ is the proton mass),  have been investigated by  
Iocco et al. (2008, hereafter I08), Freese et al. (2008b, 2009) and Spolyar et al. (2009). 
Initially, the DM pile-up is purely driven by gravitational interactions, 
but as the protostar approaches the Zero Age Main Sequence, 
DM accretion becomes dominated by the capture of WIMPs located 
in the star host halo after they scatter stellar baryons. 
As a consequence of the peculiar formation process of Pop III, 
following the smooth collapse of the gas cloud at the
very center of the DM halos hosting them,
Iocco (2008) and Freese, Spolyar \& Aguirre (2008) suggested that DM capture
 is relevant for primordial stars; however, it can be safely neglected 
once local star formation is concerned, 
as the latter takes place anywhere in galactic discs, 
and it doesn't follow from a single, centered gas collapse episode. 
Further studies (I08, Yoon, Iocco \& Akiyama 2008; 
Taoso et al. 2008) have concluded that WIMP DM capture's most 
remarkable effect is the possible increase of the stellar lifetime. 

Quite surprisingly, the early phases of the collapse have  received so far less 
attention with respect to the more advanced ones, i.e. after hydrostatic core formation. 
For example, it is still unclear if the energy 
injection following annihilations results in a nett heating or cooling of the gas. 
In fact, high energy photons and electrons heat the gas through ionizations; 
however, this heat input could be overwhelmed by the increased production 
of cooling species (as for example molecular hydrogen) stimulated by the 
larger abundances of free electrons, thus resulting in a net gas cooling. 
This, among others, is one of the aspects of the collapse of first stars in 
presence of WIMP annihilation that we would like to address here. 
We plan to do so by a set of sophisticated numerical simulations including 
all the relevant chemical reactions and cooling processes. 
A first attempt to model the effects of DMA energy input was presented in 
Ripamonti et al. (2009); this paper represents a substantial extension and 
improvement of that study.

Throughout the paper we assume the following set of cosmological
parameters: $\Omega_{\rm \Lambda}=0.76$, $\Omega_{\rm m}=0.24$,
$\Omega_{\rm b}=0.042$, $\Omega_{\rm DM} = \Omega_{\rm m} -
\Omega_{\rm b} \equiv \Omega_{\rm WIMP}= 0.198$, and $h=0.73$.

%####
\section{Method and code}\label{method}
We base our investigation on a 1-D spherically symmetric code
described by Ripamonti et al. (2002; hereafter R02). The original code, which
includes the treatment of gravitation, hydrodynamics, and especially
the chemistry and cooling of primordial gas, was originally
conceived for the study of the last phases of the collapse of a
primordial protostar (see also Omukai \& Nishi 1998); later, it was
extended in several ways (see Ripamonti 2007, Ripamonti et al. 2007a,
and RMF07).

Our simulations are based on those described in RMF07; here we list
their most important properties, especially when they differ from
RMF07:

\begin{itemize}
\item{}{A single typical halo with mass $10^6\,\msun$ virializing at
$z=20$ (virial radius $R_{\rm vir}\simeq5\times10^{20}\,\cm$) is
considered; the baryon fraction inside such halo is assumed to be
equal to the cosmological value ($\Omega_{\rm b}/\Omega_{\rm
DM}\simeq0.175$);}
\item{}{The simulations are started at $z=1000$ and involve a comoving
volume 1000 times larger than that of simulated halo; initial baryonic
density and temperature are constant in all the simulation shells, and
equal to the cosmological values;}
\item{}{Before virialization, the gravitational effects of DM are
treated as in the NFW case of RMF07: a predetermined (but
time-dependent until virialization) DM potential is added to the gas
self gravity. Such potential mimics the evolution of a halo in the
top-hat approximation: as the DM potential becomes steeper, the
(initially uniform) gas falls towards the centre of the halo, similarly to
what is predicted by theory and consistent with the results of simulations
(see e.g. Fig.~2 of Abel et al. 2002);}
\item{}{After virialization, the evolution of the previously described
artificial DM potential is stopped: its state at virialization is set
to a NFW profile (Navarro, Frenk \& White 1996) with $c=10$ and
$R_{\rm 200}=R_{\rm vir}$; this is reasonably close to the results
shown in Fig. 2 of Abel et al. (2002). The evolution of the DM profile
is also followed in response to the baryon contraction (see later) in
order to compute the DMA rate\footnote{We do not account for the
gravitational effects of the adiabatically contracted DM profile,
because adiabatic contraction is effective only when the baryonic
potential largely dominates over the DM.};}
\item{}{RMF07 investigated {\it whether} stellar formation might occur in a halo,
whereas here we investigate {\it how} it starts; for this reason,
simulations are stopped only when their computational costs become
very large (usually at number densities $n_{\rm c} \approx
10^{14}\,\percmc$);}
\item{}{The DM energy input is computed only after halo virialization,
and only in regions with high baryon density ($\rho\ge
4\times10^{-22}\,\gram\,\percmc$, i.e. $n \equiv \rho/m_{\rm p}
\gtrsim250\,\percmc$), rather than at all times and everywhere: this
is because RMF07 already showed that before virialization and at low
baryon densities the effects of DMAs are small;}
\item{}{For the purpose of evaluating the DMA rate (see below), the DM
density $\rho_{\rm DM}(r)$ is evaluated by assuming the conservation
of the so called ``adiabatic'' invariant (see Blumenthal et
al. 1986). We implement the algorithm described by Gnedin et al. (2004),
following the details in I08, and using the NFW and gas profiles described
above as initial conditions\footnote{It is to be noticed that the 
DM profile in (gravitationally) baryon dominated regions 
is eventually dictated by the amount of gas accumulated
after the contraction, see e.g. Fig 1 in SFG08.
Our conclusions, especially regarding the final phases
of the collapse, are hence almost insensitive to the initial DM profile.};}
\item{}{The specific luminosity due to DMAs is 
%###Equation
\begin{equation}
l_{\rm DM} = 
c^2\,\sigmav\,\rho_{\rm DM}^2/(m_{\rm DM}), 
\label{eq:defDMlum}
\end{equation}
%####
where $\sigmav$ is
the thermally-averaged annihilation rate, and $m_{\rm DM}$ is the WIMP
mass; in the following we adopt $\sigmav=3\times10^{-26}\,{\rm
cm^3\,s^{-1}}$, whereas we consider $m_{\rm DM}$ as a free parameter
in the range $1.78\times10^{-24}\,\gram \le m_{\rm DM} \le
1.78\times10^{-21}\,\gram$ (i.e. $1\,{\rm GeV} \le m_{\rm DM}c^2 \le 1
\,{\rm TeV}$);}
\item{}{The energy $\epsilon$ that each baryon actually absorbs from
DM annihilations (per unit time) is calculated through a
detailed radiative transfer calculation, formally identical to the one
performed for gray continuum radiation. Such calculation is based on a
constant gas opacity $\kappa=0.01\,{\rm cm^2\,g^{-1}}$ (roughly
similar to the values used by SFG08). Moreover, since it is believed
that the energy from annihilations splits roughly equally into
electrons, photons, and neutrinos, we assume that only $2/3$ of
$l_{\rm DM}$ (i.e. the fraction not going into neutrinos) can be
absorbed;}
\item{}{Similar to RMF07, $\epsilon$ can go into ionization, heating
and excitation of atoms and molecules; we employ the results of
Vald\`es \& Ferrara (2008) (see also Shull \& Van Steenberg 1985, and
Furlanetto \& Johnson Stoever 2010) to estimate how to split the
energy input into these three. Also note that the ``ionization''
component is split into ionization of H, D, He, He$^+$, and
dissociation of \HH, HD and H$_2^+$. In our ``standard''
treatment each species receives a fraction of the ionization energy
which is proportional to its total baryonic content (in number, see
RMF07 for details);
\item{}{R02 switched to equilibrium chemistry
(e.g. allowing the use of Saha equations instead of the detailed
balance ones) for shells with number densities
$n\ge10^{13.5}\,\percmc$. Here we drop this simplification since (i)
DMAs effects change the chemical evolution of the gas, and usually
delay the approach to equilibrium\footnote{Even in the few cases where
it is possible to switch to the equilibrium chemistry (e.g. the
``control'' run where we do not consider DMAs) we prefer to keep
integrating the non-equilibrium equations in order to get results
which are completely consistent with those from the other runs.}, and
(ii) we never venture to densities $n> 10^{15}\,\percmc$.}}
\end{itemize}

Given the standing ignorance on the precise detail of feedback effects
on the ionization and dissociation of atoms and molecules (especially
H$_2$), in addition to the standard, fiducial set of runs we performed
runs with either enhanced or reduced feedback, in order to bracket the
possible impact of such process. In the same way, since the opacity
$\kappa$ we employ represents only a very rough estimate, we performed
runs with either higher or lower values of $\kappa$.

\section{Results}

%%%%%%%%%%%%% TABLE 1 %%%%%%%%%%%%%
\begin{table}
\begin{center}
\caption{Properties of the runs} \leavevmode
\label{tab:runList}
\begin{tabular}{lccl}
\hline
Name & $m_{\rm DM} c^2$[GeV] & \HH fbk &
Notes\\
\hline
M1000 & 1000 & std.  &Minimal\\
M100  & 100  & std.  &Fiducial\\
M10   & 10   &  std.  &Sub-maximal\\
M1    & 1    & std.  &Maximal\\
N1000 & 1000   & off   &\\
N100  & 100    & off   &\\
N10   & 10    & off   &\\
N1    & 1      & off   &\\
E1000 & 1000   & enh.  &\\
E100  & 100   & enh.  &\\
E10   & 10     & enh.  &\\
E1    & 1      & enh.  &\\
L100  & 100  & std.  &$\kappa=0.001\,{\rm cm^2\,g^{-1}}$\\
L10   & 10   & std.  &$\kappa=0.001\,{\rm cm^2\,g^{-1}}$\\
L1    & 1    & std.  &$\kappa=0.001\,{\rm cm^2\,g^{-1}}$\\
H1000 & 1000 & std.  &$\kappa=0.1\,{\rm cm^2\,g^{-1}}$\\
H100  & 100  & std.  &$\kappa=0.1\,{\rm cm^2\,g^{-1}}$\\
H10   & 10   & std.  &$\kappa=0.1\,{\rm cm^2\,g^{-1}}$\\
NODM  & --   &  --    &Control\\
\noalign{\vspace{0.1cm}}
\hline
\end{tabular}
\end{center}
\footnotesize{The first letter of the name of a run indicates the set
to which it belongs; ``M'' refers to the main set, ``N'' to the set
without DMA feedback upon \HH formation, ``E'' to the set with
enhanced DMA feedback upon \HH formation, ``L'' to the set with low
opacity, and ``H'' to the set with high opacity.  The gray opacity is
set to $\kappa$=0.01 ${\rm cm^2\,g^{-1}}$ for all of the runs
presented in this Table, except for runs in the ``L'' and ``H'' sets.
The NODM run assumes no energy input from DMAs.}

\end{table}
%%%%%%%%%% TABLE 1 - END %%%%%%%%%%

We test the effects of DMAs varying different sets of parameters: {\it
(i)} the normalization of the DMA rate, which is regulated by the
ratio $\sigmav$/$m_{\rm DM}$; {\it (ii)} the feedback strength on
chemistry; and {\it (iii)} the ``gray'' gas opacity $\kappa$.  We
anticipate that the strength of feedback on chemistry has little
impact on the overall results, and that the effects of a variation in
$\kappa$ are somewhat similar to a variation of the same amount in
$\sigmav$/$m_{\rm DM}$.  We will discuss the dependence on these two
latter parameters in Section \ref{Sec:opacFeed}.

Here we start by introducing in more details the physics of first star
formation in presence of DMAs for our fiducial models (``M'' labeled
runs, see Table \ref{tab:runList} for details).  It is worth
anticipating that the effects of DM become more relevant (and they
become efficient earlier during the collapse) for higher
self-annihilation cross sections/lower DM masses (see
eq.~\ref{eq:defDMlum}). In what follows, we always adopt
$\sigmav$=3$\times$10$^{-26}$ cm$^3$/s, and vary the particle mass
$m_{DM}$. Given the degeneracy in the DMA energy injection term, the
results can be interpreted at fixed mass and correspondingly varying
the self-annihilation rate.

%%%%%%%%%%%%% FIGURE 1 %%%%%%%%%%%%%
\begin{figure}
\centering
\includegraphics[angle=0,width=0.48\textwidth]{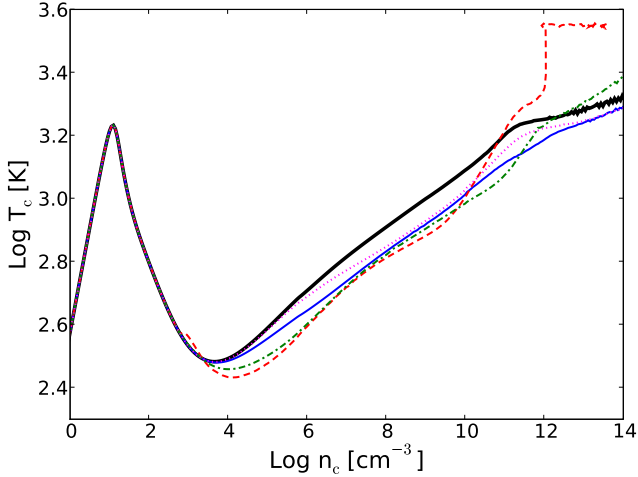}
\caption{Evolution of the temperature $T_{\rm c}$ of the central shell
of our simulation (whose mass is $\simeq2\times10^{-4}\,\msun$) as a
function of its baryon number density $n_{\rm c}$, starting slightly
before virialization. The thick solid line shows the evolution in the
control run with no DM energy input (NODM).  The thin solid (blue)
corresponds to the fiducial model (run M100); the dashed (red) to the
maximal model (M1); the dot-dashed (green) to the sub-maximal model
(M10); and the thin dotted (magenta) to the minimal model (M1000).}
\label{fig:plot_n_T}
\end{figure}
%%%%%%%%%% FIGURE 1 - END %%%%%%%%%%

Figure~\ref{fig:plot_n_T} shows the evolution of the temperature
$T_{\rm c}$ of the innermost shell as a function of the density of the same
central shell of the simulated objects ($n_{\rm c}$), for
the five most representative models, M100, M10, M1, M1000,
and NODM (see Table~\ref{tab:runList}).

As expected, DMA effects, which can be quantified by the deviations
from the thick solid line, are most prominent in runs with low mass
particles $m_{\rm DM} c^2 = 1\,{\rm GeV}$ (M1 run, corresponding to
the maximal DMA energy injection rate), and become very limited in the
case of DM particles with high mass $m_{\rm DM} c^2 = 1\,{\rm TeV}$
(M1000 run, minimal DMA energy injection rate); we will comment
extensively on such dependence later in this Section.  In the
following we will refer to the ``fiducial'' run (M100; where the
choice of parameters is quite standard), to the ``minimal'' run
(M1000; where the high value of $m_{\rm DM}$ reduces the energy
injection from DMAs, providing a check on the level where its effects
become negligible), to the ``maximal'' run (M1; where the parameters
are chosen in a way which maximizes the energy injection from DMAs and
their effects upon the gas evolution), and to the ``sub-maximal'' run,
M10.  The maximal model, M1, is likely to represent a sort of upper
limit on the effects of DMAs on the formation of primordial stars. In
fact, the DM parameters corresponding to this run
($\sigmav$=3$\times$10$^{-26}\,{\rm cm^3\,s^{-1}}$, $m_{\rm DM} c^2$=1
GeV) are currently severely disfavored by multimessenger constraints
on DM, in particular by (astrophysical) parameter-independent ones
(e.g. Galli et al. 2009; Cirelli, Iocco and Panci 2009).

%#####
\subsection{The indirect feedback phase}
\label{sec:indirfeed}
%#####

%%%%%%%%%%%%% FIGURE 2 %%%%%%%%%%%%%
\begin{figure}
\centering
\includegraphics[angle=0,width=0.48\textwidth]{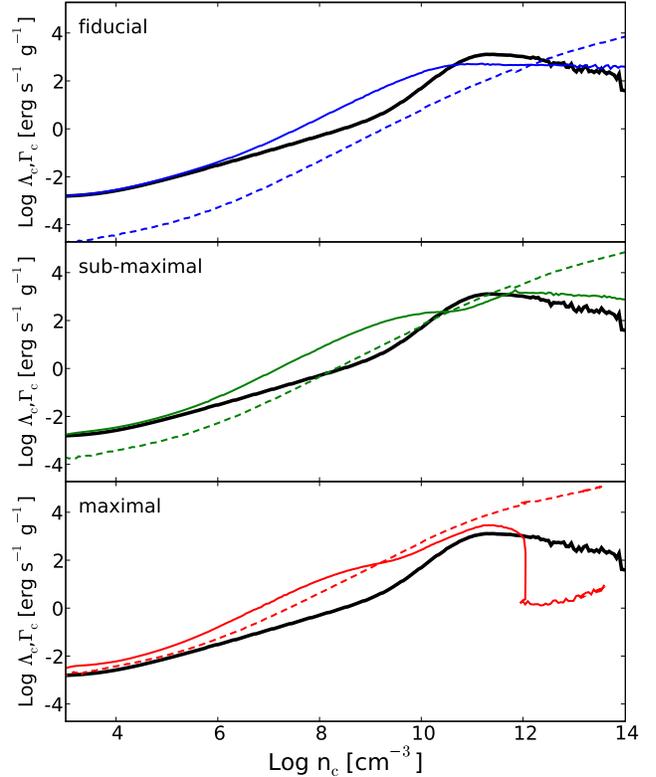}
\caption{Evolution of the \HH cooling rate and of the
rate of DMA energy input in the central
shell of our simulations, as a function of its number density. In each
panel the \HH cooling rate (thin solid line) and the DMA energy input
(dashed line) for one of our runs (from top to bottom: fiducial, sub-maximal, 
and maximal runs) are compared to the \HH cooling rate in the control NODM run 
(thick solid line). We note that the actual heating rate from DMAs is about 
$1/3$ of what is shown here, since the total DMA energy input also includes 
energy actually going into excitations and ionizations (see text for details).}
\label{fig:plot_n_LambdaH2_DMinput}
\end{figure}
%%%%%%%%%% FIGURE 2 - END %%%%%%%%%%

Figure~\ref{fig:plot_n_LambdaH2_DMinput} compares the total energy
input from DMAs into the gas (which needs to be splitted into
ionizations, heating and excitations) to the cooling rate due to
\HH. This comparison largely oversimplifies the thermal balance (see
next subsection); however, it makes clear that, even in the maximal
case, the heating from DMAs remains a relatively small fraction of the
\HH cooling until $n_{\rm c}$ becomes $\gtrsim 10^9\,\percmc$. At
earlier stages (i.e. lower densities) the effects of DMAs are more
subtle than what could be simply inferred by the direct heating, which
would not be able to significantly change the temperature of the gas.
In fact, it is quite remarkable that in the $10^3\,\percmc \lesssim
n_{\rm c} \lesssim 10^{11}\,\percmc$ range, the {\it additional} energy
injection from DMAs results in a {\it decrease} of the central
temperature (see also Fig.~\ref{fig:plot_n_T}).

Such counter-intuitive temperature decrease is due to indirect
effects of DMAs, and in particular to their feedback on the chemistry;
this remarkable property clearly emerges upon inspection of the curves in
Fig.~\ref{fig:plot_n_LambdaH2_DMinput} showing that \HH cooling
(thin solid line) is stronger when DMAs are included 
(at least for $n_{\rm c} \lesssim10^{10}\,\percmc$).

\begin{figure}
\centering
\includegraphics[angle=0,width=0.48\textwidth]{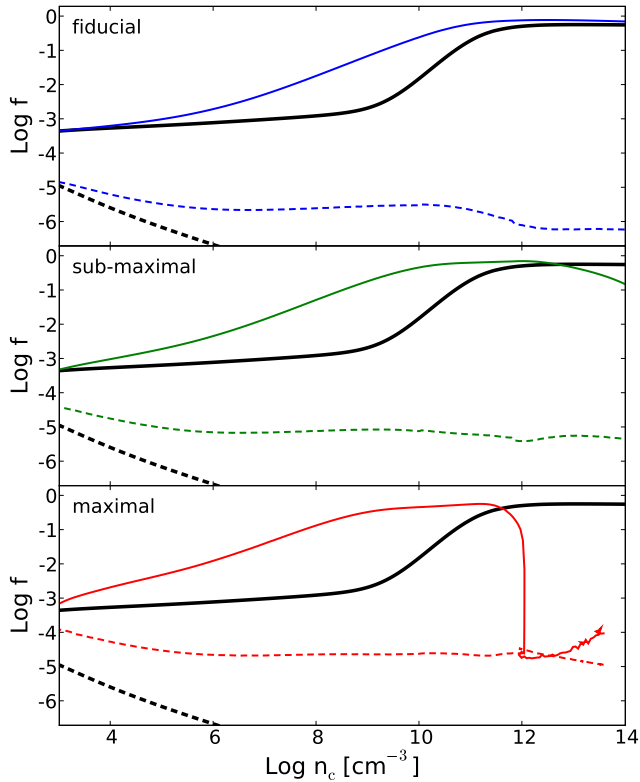}
\caption{Evolution of the \HH and $e^-$ fractions in the central shell
of our simulation as a function of its baryon number density. In each
panel the \HH fraction (thin solid line) and $e^-$ fraction (thin
dashed line) from one of our runs (from top to bottom: fiducial, 
sub-maximal, maximal runs) are compared to the same quantities 
from the control NODM run (thick solid line for \HH and thick dashed 
line for $e^-$).}
\label{fig:plot_n_f}
\end{figure}

%Of course, as we have already noted when discussing Fig.~\ref{fig:plot_n_f}, the previous
%result can bee seen as an alternative way to re-state that presence 
%of DMAs leads to a faster increase in the \HH fraction. 
In fact, as it can be inferred from Fig.~\ref{fig:plot_n_f}, 
DMA {\it ionization} effects keep the gas more ionized than in the NODM 
case (i.e. $e^-$ abundance $10^{-6}$--$10^{-4}$, rather than $\ll
10^{-6}$). In turn, this relatively high ionization level favors the
formation of \HH through the reaction chain
\begin{eqnarray}
{\rm H} + e^- &\rightarrow& {\rm H}^- + \gamma\\
{\rm H}^- + {\rm H} &\rightarrow& {\rm H}_2 + e^-,
\end{eqnarray}
which is the main formation route for \HH in moderately dense,
dust-free gas, and where the electrons act as catalysts (see
e.g. Galli \& Palla 1998; Glover \& Jappsen 2007; Stasielak, Biermann
\& Kusenko 2007).

%%%%%%%%%%%%% FIGURE 3 %%%%%%%%%%%%%
%%%%%%%%%% FIGURE 3 - END %%%%%%%%%%

As Ripamonti et al. (2009) remarked, the increase in the \HH fraction
can amount to 2-3 orders of magnitude, which apparently might imply same
order-of-magnitude differences in the cooling rate, and large
differences also in the gas temperature. However, while the increase
in the cooling rate is quite large (though smaller than what is
naively expected from the difference in \HH abundance), the corresponding
reduction of the gas temperature is much smaller, of the order of 30 per
cent (see Fig.\ref{fig:plot_n_T}). This is due to the very strong temperature 
dependence of \HH cooling, especially below $\sim 500\,\kelvin$ (the temperature
corresponding to the transition from the fundamental state to the
lowest rotationally excited level): in fact, in this regime a moderate
reduction in temperature results in a much larger decrease in the \HH
cooling rate, which in turn slows down the temperature decrease.

%%%%%%%%%%%%% FIGURE 4 %%%%%%%%%%%%%
\begin{figure}
\centering
\includegraphics[angle=0,width=0.48\textwidth]{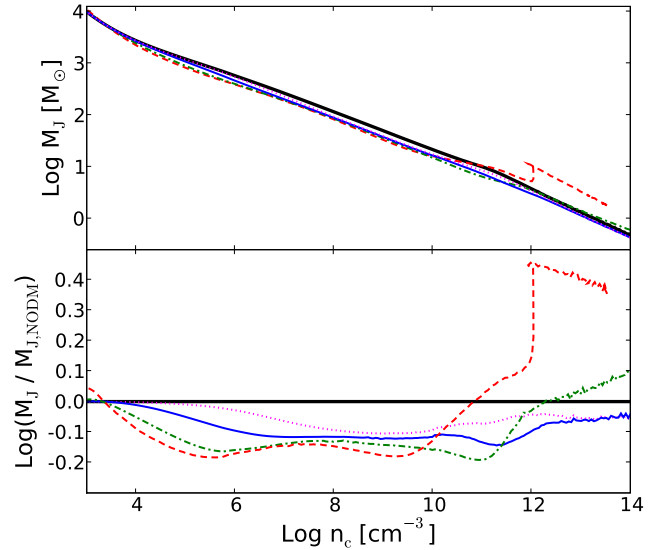}
\caption{Evolution of the Jeans mass calculated using the density and
  temperature in the central shell of our simulation, as a function of
  its number density. Lines refer to the minimal (M1000; dotted),
  fiducial (M100; thin solid), sub-maximal (M10; dot-dashed), maximal
  (M1; dashed) models, and to the control run (NODM; thick solid). The
  top panel shows the absolute value of $M_{\rm J}$, whereas the
  bottom panel compares the various models by showing the ratio of
  $M_{\rm J}$ in a given model to $M_{\rm J,NODM}$ in the NODM run.}
\label{fig:plot_n_mjeans}
\end{figure}
%%%%%%%%%% FIGURE 4 - END %%%%%%%%%%

As a result, since the Jeans mass sensitivity to the temperature is quite mild,
\begin{equation}
M_{\rm J}(n,T) \simeq
50{\rm M_\odot} \left({T\over{1\,\kelvin}}\right)^{3/2}
\left({n\over{1\,{\rm cm^{-3}}}}\right)^{-1/2},
\end{equation}
DMAs might reduce the Jeans mass scale but only by a factor $\lesssim 2$ (see Fig. \ref{fig:plot_n_mjeans}).

%#####
\subsection{The direct feedback phase}
\label{Sec:Dirfeed}
%#####

%%%%%%%%%%%%% TABLE 2 %%%%%%%%%%%%%
\begin{table}
\begin{center}
\caption{Density and temperatures where the total energy input from DMAs
overcomes \HH cooling.} \leavevmode
\label{tab:criticalPoints}
\begin{tabular}{ccc}
\hline
$m_{\rm DM} c^2$ & $n_{\rm crit}$[$\percmc$] & $T_{\rm crit}$[$K$]\\
\hline
1 GeV$^{(a)}$    & $1.5\times10^{9}$ ($\sim10^9$)$^{(b)}$ &
                   780 ($\sim$800)$^{(b)}$\\
10 GeV$^{(a)}$   & $4.0\times10^{10}$ ($\sim 10^{11}$)$^{(b)}$         &
                   1070 ($\sim 1000$)$^{(b)}$\\
100 GeV$^{(a)}$  & $1.3\times10^{12}$ ($\sim10^{13}$)$^{(b)}$     &
                   1580 ($\sim$1300)$^{(b)}$\\
1000 GeV$^{(a)}$ & $8.0\times10^{12}$ ($\sim 10^{14}$)$^{(b)}$         &
                   1740 ($\sim 1400$)$^{(b)}$\\
\noalign{\vspace{0.1cm}}
\hline
\end{tabular}
\end{center}
\footnotesize{Notes: (a) The quantities in this table are almost
independent of the assumptions about the strength of \HH feedback, so we do not
distinguish among runs with the same $m_{\rm DM}$. (b) Numbers in
brackets indicate the results which can be estimated from Fig. 2 of
SFG08, intersecting the Yoshida et al. (2006) curve with the curves for
various DM masses; for the SFG08 1 GeV mass case we adopt the 100\%
\HH line.}
\end{table}
%%%%%%%%%% TABLE 2 - END %%%%%%%%%%

Figure \ref{fig:plot_n_LambdaH2_DMinput} shows that when the density
increases above $n_{\rm c} \sim 10^{12}\, \percmc$ ($\sim 10^{9}\,
\percmc$) for the fiducial (maximal) run, the energy input from DMAs
finally overcomes the \HH cooling (see also Table
\ref{tab:criticalPoints} for details and a comparison with previous
literature).  This marks the beginning of what we call the direct
feedback phase of the collapse, since the DMAs {\it direct} effects
(especially the heating) finally start to dominate both \HH cooling
and the more subtle DMAs indirect effects discussed above. We will
refer to this condition (equality of DMA heating and \HH cooling
terms) as to the ``critical point''.

The study of this phase is particularly interesting because when the DMA
heating starts to compensate the radiative cooling, the protostar
becomes unable to shed away its gravitational energy; it had been
previously proposed (see SFG08) that this could induce the
stop of the collapse, and the formation of a new type of stable
astrophysical object powered by DMAs.

The evolution of all the quantities beyond the critical point, 
shown in  Figures~\ref{fig:plot_n_T} and~\ref{fig:plot_n_LambdaH2_DMinput},
imply the existence of a dynamical evolution of the system
also after the onset of the direct feedback phase -and
hence no stalling of the object-
even if there are important differences among the
various runs. 

In the fiducial run, the evolution of the
central shell proceeds relatively undisturbed. For 
$n_{\rm c} \le 10^{14}\,\gram\,\percmc$ the central temperature remains below
the one of the control run, even if the gap tends to close
(see Fig.~\ref{fig:plot_n_T}). Conversely, in the maximal run there is a 
substantial steepening of the evolution in the $n_{\rm c}-T_{\rm c}$ plane, and the central
temperature overcomes the value of the control run at $n_{\rm c} \sim 10^{11} \,\percmc$. 
The temperature increase becomes particularly dramatic, with a sudden rise 
by a factor $\sim 2$ when the central density reaches $n_{\rm c} \sim 10^{12}\,\percmc$.

In order to understand this phase, we need to describe the heating and
cooling processes in detail. Fig.~\ref{fig:plot_n_heatcool} shows the
contributions from the main heating and cooling mechanisms during the
contraction of the central shell, for our reference runs.

%####
\subsubsection{Details of chemical cooling}
\label{Sec:detchemcool}
%####

%%%%%%%%%%%%% FIGURE 5 %%%%%%%%%%%%%
\begin{figure}
\centering
\includegraphics[angle=0,width=0.48\textwidth]{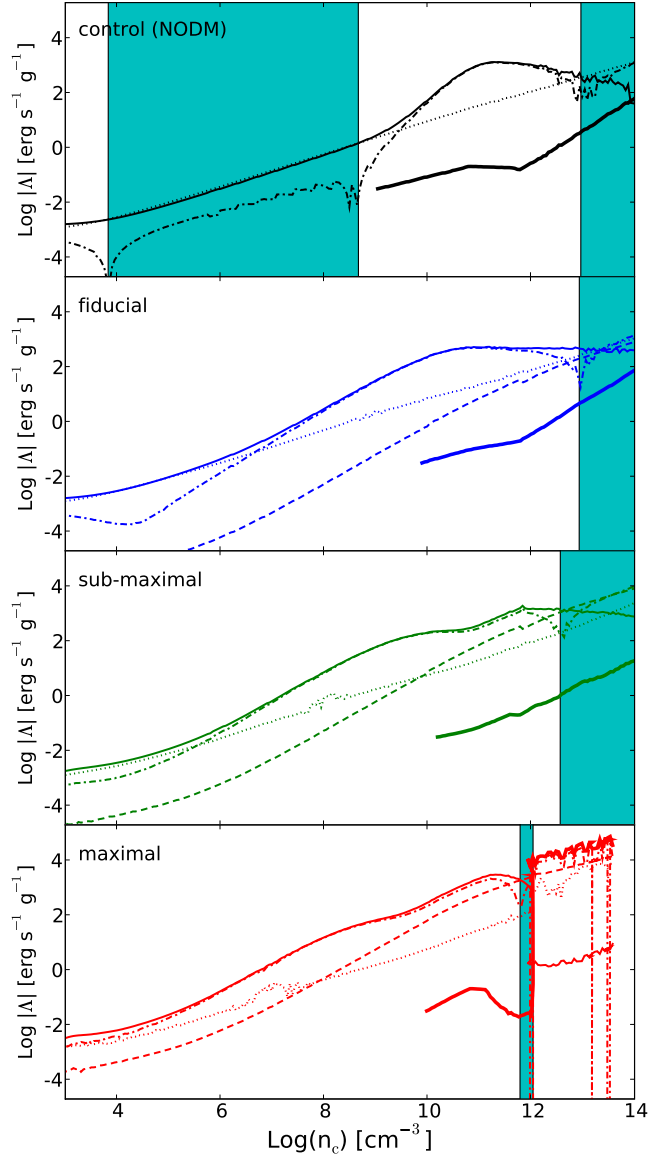}
\caption{Evolution of the main heating and cooling mechanisms in the
  central shell of our simulations, as a function of its number
  density $n_{\rm c}$. The panels refer to four different runs: from
  top to bottom, we show results for the control (without DMAs),
  fiducial, sub-maximal, maximal runs. Each panel shows \HH line
  cooling (thin solid line), continuum cooling (thick solid line), DMA
  heating (dashed line), adiabatic heating (dotted line), and chemical
  heating or cooling (dot-dashed line). We note that i) the continuum
  cooling line can be plotted only when $T_{\rm c}\ge1000\,\kelvin$
  (corresponding to $n_{\rm c}\gtrsim10^{9}-10^{10}\,\percmc$),
  however it is negligible at lower temperatures; ii) the DMA heating
  is different from the DMA energy input shown in
  Fig.~\ref{fig:plot_n_LambdaH2_DMinput}, since it does not include
  the ionization and excitation components; iii) the chemical term
  does not have a constant sign: shaded regions indicate the density
  ranges where it corresponds to a cooling term, whereas the ranges where
  it corresponds to a heating term are not shaded.}
\label{fig:plot_n_heatcool}
\end{figure}
%%%%%%%%%% FIGURE 5 - END %%%%%%%%%%

Figure \ref{fig:plot_n_heatcool} shows the large importance
of chemical heating/cooling, i.e. of the energy which is released/absorbed 
as a consequence of chemical reactions, and in particular 
(in this regime, at least) of the formation/dissociation
of \HH molecules\footnote{The formation of a \HH molecule releases
its binding energy ($4.48\,{\rm eV}$); at the densities we are
considering here, most of this energy is eventually converted into
thermal energy of the gas (see e.g. Hollenbach \& McKee 1979); on the
other hand, the dissociation of \HH through collisional reactions
absorbs the same amount of thermal energy from the gas.}: in every run 
there is a phase where \HH formation is the main heating
mechanism. In the control run, chemical heating dominates over the adiabatic 
heating in the $10^{9.5}\,\percmc \lesssim n_{\rm c} \lesssim 10^{12.5}\,\percmc$ 
density range, which coincides with the regime where 3-body reactions (see Palla, Salpeter
\& Stahler 1983) turn most of the Hydrogen into molecular form. In the
runs with DMAs such a phase is even more extended, as the DMA
feedback on the chemistry anticipates the epoch of rapid \HH formation: in the maximal
run, chemical heating becomes dominant already at $n_{\rm c}\gtrsim10^{3.5}\,\percmc$.

It is remarkable that the chemical term can contribute both to heating
and to cooling. It was already known (see e.g. Omukai \& Nishi 1998;
R02) that because of this property, it tends to act
as a ``thermostat'' which stabilizes the evolution of the collapse,
even if it cannot always prevent sudden transitions.

The evolution of the maximal and sub-maximal models are particularly
significant in this respect. In these models the \HH fraction peaks
(at a level $\gtrsim 0.8$; see Fig.~\ref{fig:plot_n_f}) when
$n_{\rm c}\simeq10^{11}$--$10^{12}\,\percmc$: in such conditions 3-body \HH
formation is slowed down by the relative scarcity of atomic H left, and
must compete with the increasingly large number of dissociations due
to the ionization component of the DMA energy input\footnote{It is
important to note that the DMA energy injection which goes into
ionizations actually contributes to the chemical heating. For example,
if the number density of \HH molecules remains constant because the
rate of formation (through 3-body reactions) is compensated by the
rate of dissociations induced by the DMAs, the chemical heating term 
will be positive. In fact, the energy released by \HH formation ends up in the form of thermal
energy, whereas the energy for the dissociations is {\it not} taken
away from the gas thermal energy, as it is provided by DMAs.
%We further note that the lack of a temperature discontinuity at
%$n_{\rm c}\sim10^{12}\,\percmc$ in the evolution of the N1 model of
%Fig.~\ref{fig:plot_n_T} is explained by the fact that when \HH
%dissociation feedback is switched off, chemical heating is much
%less affected by the DMA energy input.
}.

When the molecular fraction starts to go down, the cooling rate 
must follow, because it is dominated by \HH and the number of
molecules is being reduced; but so does the heating, because its
dominant component (chemical heating) is reduced by the energy lost
in molecular dissociations. The fact that both the heating and the
cooling mechanisms follow the same trend prevents a significant
steepening of the temperature evolution and the consequent stop in the
protostellar collapse. This is because the ``excess'' energy which is
not radiated away is mainly used to dissociate \HH.

However, this cannot last indefinitely, as the reservoir of \HH is
finite, and the DMA energy injection increases with increasing
density. In the case of the maximal run, where the DMA energy
injection is quite high, \HH is rapidly exhausted. At $n_{\rm
c}\simeq10^{12}\,\percmc$ there is essentially no \HH left, which also
implies almost no radiative cooling. Then, the temperature goes up
very fast (see Fig.~\ref{fig:plot_n_T}), developing a pressure
gradient which is able to momentarily stop the collapse of the
innermost shell (actually, there is even a very brief re-expansion
episode- more details are discussed in Section
\ref{sec:EvolSpatProf}); however, the increase in temperature also
results in a very strong increase in the continuum radiation which is
produced by collision-induced emission, H$^-$, and atomic H: continuum
radiation increases by some 6 orders of magnitude (cfr. Lenzuni,
Chernoff \& Salpeter 1991; Marigo \& Aringer 2009) and replaces \HH
line radiation as the main cooling mechanism for the gas. The collapse
is then able to resume, even if further investigations are needed in
order to understand its details beyond $n_{\rm c}\approx
10^{13}\,\percmc$.  At present in fact, computational costs become
almost unaffordable beyond $n_{\rm c}\approx 10^{13}\,\percmc$ because
of the strong sensitivity of cooling rate on temperature. In spite of
these limitations we consider the conclusions based on these runs very
robust as we have pushed the evolution to densities exceeding $10^4$
times the ''critical" density at which the \HH cooling becomes
comparable to the DMA energy transferred to baryons.

Things are quite different in the case of the sub-maximal run: here a
moderate \HH dissociation rate can compensate both the heating and the
ionization parts of the (relatively small) DMA energy input for a
significant span. In fact, at $n_{\rm c}=10^{14}\,\percmc$ (i.e. at a
density $\sim 100$ times larger than the density at the onset of \HH
dissociation, and $\sim 1000$ times larger than the density at the ``critical point'') 
the \HH fraction is still $\gtrsim 0.1$: it is still
possible that \HH exhaustion (which we expect to happen at
$n_{\rm c}\sim10^{15}\,\percmc$) is accompanied by a transition similar to
the one we discussed for the maximal case; however, such transition is
likely to involve a smaller adjustment in temperature (say, an
increase by a factor $\sim 1.5$ rather than $\sim 2$) because the
required increase in the continuum cooling is much smaller (somewhat
less than 3 orders of magnitude rather than 6, as can be inferred from
Fig.~\ref{fig:plot_n_heatcool}) than what was required in the maximal
case.

The fiducial and minimal runs exhibit trends which are somewhat
similar to the sub-maximal run: the peak in the \HH  abundance at
$n_{\rm c}\sim10^{12}\,\percmc$ is followed by an extremely slow decline,
during which the adiabatic heating is often as important as the DMA
heating: then, it is quite likely that the transition from \HH line
cooling to continuum cooling is as smooth as in the scenario without
DMAs. In fact, it should be noted that at $n_{\rm c}=10^{14}\,\percmc$ the
continuum cooling amounts already to about 10 per cent of the total
heating, and it is increasing faster than any heating mechanism (see
Fig.~\ref{fig:plot_n_heatcool}).

%#####
\subsection{Sensitivity to feedback and opacity}
\label{Sec:opacFeed}
%####

%%%%%%%%%%%%% FIGURE  6 %%%%%%%%%%%%%
\begin{figure}
\centering
\includegraphics[angle=0,width=0.48\textwidth]{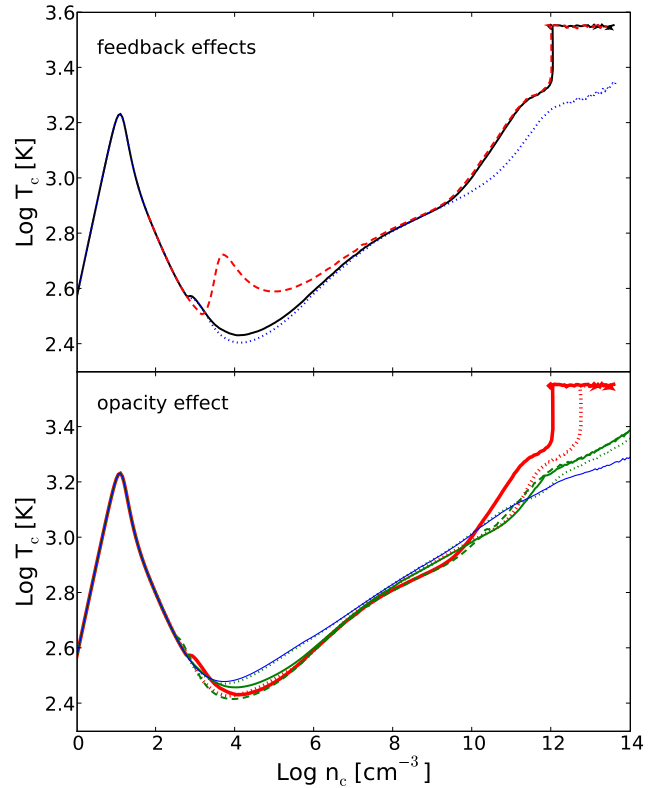}
\caption{Effects of \HH feedback schemes (upper panel) and opacity
(lower panel): in the upper panel the solid (black) represents the
maximal run M1, the dashed (red) line is the E1 run (case with
enhanced \HH feedback) and dotted (blue) line is the N1 run (no \HH
feedback). In the lower panel thick (red) lines represent the M1
(solid) and L1(dotted) runs, the intermediate (green) lines represent
the the M10 (solid) the L10 (dotted) and H10 (dashed), finally, the
thin (blue) line is our fiducial M100 run. See text for details and
Table \ref{tab:runList} for run coding.}
\label{fig:plot_T_fbk_kappa}
\end{figure}
%%%%%%%%%% FIGURE 6 - END %%%%%%%%%%

In order to check the effect of our assumptions about \HH feedback (in
particular, destruction of \HH molecules by the energy injected by
DMAs into the gas) we have run two extra sets of simulations: the
first (label: E) in which \HH feedback is enhanced, and the second
(label: N) with DM-induced \HH dissociations switched off.  In the
E-models, we have assumed that {\it all} DMA energy input which goes
into excitations is in the form of photons in the Lyman-Werner band
($11.2\,{\rm eV}\le h\nu \le 13.6\,{\rm eV}$), and that each of these
photons dissociates one \HH molecule\footnote{In practice, we assume a
Lyman-Werner \HH dissociation rate (per unit volume) equal to $n
\epsilon f_{\rm exc} / \bar{E}_{\rm LW}$, where $n$ is the baryon
number density, $\epsilon$ is the energy input per baryon due to DMAs,
$f_{\rm exc}$ is the fraction of this energy which goes into
excitations (i.e. into photons with $10.2\,{\rm eV}\le h\nu \le
13.6\,{\rm eV}$), and $\bar{E}_{\rm LW}\simeq 12\,{\rm eV}$ is the
average energy of a Lyman-Werner photon. Such rate is added to the one
produced by the fraction the DMA energy input which goes into
ionizations.}.  This clearly represents an upper limit on the feedback
upon \HH, as in reality about half of the photons going into
dissociations has energy below the lower limit of the Lyman-Werner
band; furthermore, in low density regions a significant fraction of
Lyman-Werner photons can escape.

As it can easily be noticed from the upper panel of Figure
\ref{fig:plot_T_fbk_kappa}, (showing the evolution of models M1, E1
and L1 in the $n_{\rm c}-T_{\rm c}$ plane) switching off the \HH
feedback results in a sensible change of the temperature only in the
final stages of collapse (because the absence of DMA-induced \HH
dissociations strongly reduces the heating); on the other hand,
strengthening the feedback results in appreciable modification of the
temperature of the cloud in earlier stages (because it reduces the \HH
abundance), but the difference reduces drastically in later phases
(when the direct effects of DMAs dominate the ones on chemistry).
We also note that in models with $m_{\rm DM} c^2 \ge 10\,{\rm GeV}$
the differences induced by the treatment of feedback are much smaller
than in the models shown in Fig.~\ref{fig:plot_T_fbk_kappa} (with
$m_{\rm DM} c^2 = 1\,{\rm GeV}$).

% The enhanced feedback model, E1, significantly differs from run M1
% at intermediate densities ($10^3\,\percmc\lesssim n_{\rm c} \lesssim 10^7\,\percmc$), but the difference 
% is strongly reduced (although it never disappears) at higher densities. Instead,
% Runs N1 and M1 deviate 
% only at high density ($n_{\rm c}\gtrsim 10^{12}\,\percmc$). The reasons of these differences will be addressed later; 
% here, we simply point out that they are quite limited, and we will hence focus on the main set of runs (models M) 
% with intermediate feedback values. 

We have also studied the effects of varying the gray opacity $\kappa$,
by running sets of simulations in which we have increased/decreased
the fiducial value $\kappa$=0.01$\,{\rm cm^2\,g^{-1}}$ by one order of
magnitude.  At low and intermediate densities the effects on the
evolution of the cloud are very similar to those of a
decrease/increase of the particle mass $m_{\rm DM}$; whereas models
with the same $m_{\rm DM}$ tend to converge at high densities. In
fact, an opacity variation affects sensibly the properties of the gas
only when it is of ``optically thin'' to the energy produced by DMAs:
in later stages of the collapse, (i.e. in regions of the halo where
the gas is ``optically thick'' to the DMA energy), the energy
absorption does not depend on $\kappa$.  This can be
appreciated from the lower panel of Figure \ref{fig:plot_T_fbk_kappa},
where it is clear that the modified opacity models (H, L labeled)
overplot to the corresponding standard (M labeled) models at higher
central densities.

We can definitely infer that ``astrophysical'' (and ``numerical'')
parameter dependence is mostly degenerate with DM parameters,
and in any case not drastically affecting the qualitative picture we
have drawn so far.

%####
\subsection{The evolution of spatial profiles}
\label{sec:EvolSpatProf}
%#####

\subsubsection{The fiducial run}

%%%%%%%%%%%%% FIGURE 7 %%%%%%%%%%%%%
\begin{figure}
\centering
\includegraphics[angle=0,width=0.48\textwidth]{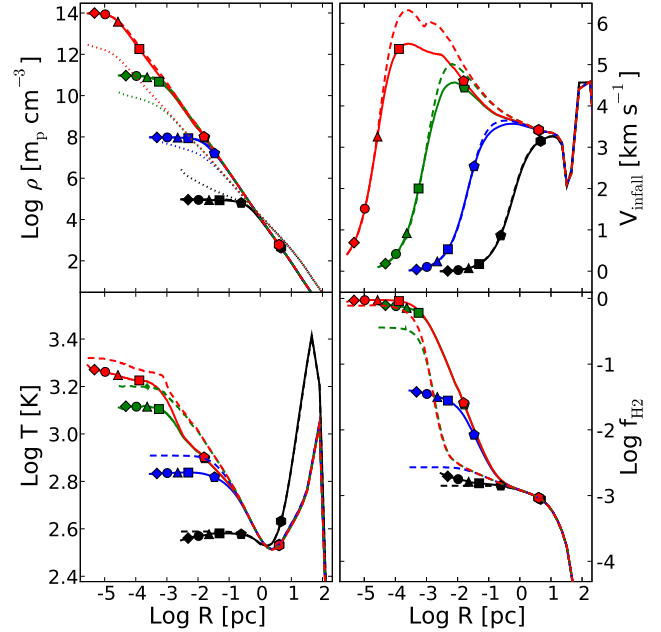}
\caption{
Evolution of DM and gas profile (top left), infall velocity (top right),
temperature (bottom left), and \HH fraction (bottom right) 
during the collapse, comparing runs M100 (fiducial) and NODM (control).
Solid lines refer to gas quantities in the M100 case,
dashed lines to gas quantities in the NODM control run.
Different solid lines refer to
central shell gas densities $n_{\rm c}$=10$^5$ (black), 
10$^8$ (blue),  10$^{11}$ (green), 
10$^{14}\,\percmc$ (red); the different
stages can be easily distinguished through the position of their
innermost point, as the lines referring to high-density stages extend
to lower radii. 
In the upper left panel, the dotted lines refer to DM densities
in the M100 case (note that the DM
density profile for the $n_{\rm c}=10^5\,\percmc$ stage is very
similar to the NFW profile which is assumed to form immediately after
virialization).
The geometrical markers indicate the position
enclosing baryonic masses of 10$^4$M$_\odot$ (hexagons),
10$^2$M$_\odot$ (pentagons), 1M$_\odot$ (squares), 10$^{-1}$M$_\odot$
(triangles), 10$^{-2}$M$_\odot$ (circles), and 10$^{-3}$M$_\odot$
(diamonds); see text for comments.}
\label{fig:profilo_r}
\end{figure}
%%%%%%%%%% FIGURE 7 - END %%%%%%%%%%

Figure~\ref{fig:profilo_r} shows the spatial profile of gas properties
at four different stages (corresponding to central shell gas densities
$n_{\rm c}=10^5$, $10^8$, $10^{11}$, and $10^{14}\,\percmc$) in the
M100 fiducial case and, for comparison, in the NODM control run. The
top-left panel shows also the DM density in the case of M100 run
at the same stages (DM density stays constant and equal to the initial profile in the NODM control
run). 

The first important remark to file is that the gas density profiles are
virtually unchanged by the presence of DMAs: such profiles are
indistinguishable until central gas densities of $n_{\rm
c}\sim$10$^8\percmc$ are reached, and even then the discrepancies are
of minor entity (e.g., the slope outside the core becomes $\simeq
2.15$ rather than $\simeq 2.20$). We can conclude that DMAs do not
alter the self-similar phase of the collapse (Larson 1969, Penston
1969), at least for run M100 and for $n_{\rm c}\le 10^{14}\,\percmc$.
The evolution of the DM density profile is in good agreement with
previous studies, see for instance Fig. 1 in SFG08, and Fig. 1 in I08.

It is also relevant to notice that (as could be expected) the
differences in infall velocity, temperature profile and \HH fraction
affect only the central $\sim 10^4\,\msun$ of gas (i.e. the gas within
the hexagonal marks). Actually, relevant (although yet not dramatic)
differences are limited to the inner $\sim 10^2\,\msun$ (region within
the pentagonal marks): the overall properties of the $10^6\,\msun$
halo are conserved in presence of DMAs.

The lower panels visualize the changes in temperature
(lower left) following the modification of \HH fraction (lower right)
which we have extensively commented in the previous sections.
Here we only note
that the molecular core is significantly more extended than in the
control run: the region where \HH is the dominant chemical species
encloses $\sim 5\,\msun$, rather than $\sim 1\,\msun$; furthermore,
the fall in \HH abundance is less steep than in the case without
DMAs.

The upper right panel is somewhat intriguing, as it gives a glimpse of
the modification of the dynamical properties of the collapse: while
there are very little changes in the central region (say, within the
square marks, corresponding to a $1\,\msun$ enclosed mass) and in the
outskirts of the halo (at enclosed masses $\gtrsim 10^3\,\msun$,
i.e. outside a point roughly mid-way between the pentagonal and
hexagonal marks), the infall velocity of the intermediate-mass gas
shells is slowed down. The reduced temperature of the cloud (due to
the more efficient \HH cooling which we discussed in the previous
sections; see also the bottom-right panel of Fig.~\ref{fig:profilo_r})
is likely responsible for this change. In fact, it is well known (see
e.g.  Stahler, Palla \& Salpeter 1986, and references therein) that
during the self-similar evolution, the mass infall rate $\rho(r) v_{\rm
infall}(r)$ in the region where the density falls as a power law with
index $\sim 2$ should roughly scale with $c_{\rm s}^3/G$, where
$c_{\rm s}\propto T^{1/2}$ is the isothermal sound speed. Since
$\rho(r)$ is practically the same in the M100 and the NODM run, the
reduced temperature should result in a reduced infall velocity,
similar to what we obtain.
%Such modification is however not
%dramatic, amounting to less than one order of magnitude, only for a
%short time, and only for few shells.
%The behaviour of the infall
%velocity discrepancy (M100 vs NODM runs) is in fact non-trivial, and
%worth commenting: as it can be seen from the upper right panel, the
%infall velocity, after slightly diverging at enclosed gas masses of
%10$^3$M$_\odot$ reconverges toward the NODM case at the innermost
%shells.

It is worth noting that on the contrary, the collapse of the
innermost gas shell is {\it accelerated}, albeit very
slightly, by the presence of DM, until central densities of
10$^{12}\percmc$(/10$^{11}\percmc$) for the M100(/M1) case, see Table
\ref{tab:collapsetime}.  However, such infall time alteration are of
very small entity, and in any case negligible with respect to the
total collapse time -- of the order of $3\times10^6\,{\rm yr}$.

\begin{table}
\begin{center}
\caption{Collapse times (yr) of the innermost gas shell} \leavevmode
\label{tab:collapsetime}
\begin{tabular}{lccc}
\hline
Log ($n_{\rm c}$[$\percmc$])   &    NODM  &   M100 &    M1\\
\hline
2--3    &    2.117$\times$10$^6$     &          2.116$\times$10$^6$    &          2.107$\times$10$^6$\\
3--4     &   8.382$\times$10$^5$       &        8.326$\times$10$^5$     &         8.077$\times$10$^5$\\
4--5      &  3.204$\times$10$^5$      &         3.200$\times$10$^5$    &          3.401$\times$10$^5$\\
5--6    &    1.158$\times$10$^5$      &         1.177$\times$10$^5$    &          1.190$\times$10$^5$\\
6--7    &    3.930$\times$10$^4$      &         4.026$\times$10$^4$    &          3.895$\times$10$^4$\\
7--8   &     1.261$\times$10$^4$      &         1.250$\times$10$^4$    &          1.221$\times$10$^4$\\
8--9   &     3.927$\times$10$^3$      &         3.757$\times$10$^3$    &          3.544$\times$10$^3$\\
9--10   &    1.210$\times$10$^3$    &           1.118$\times$10$^3$   &           1.072$\times$10$^3$\\
10--11  &     366.4      &           333.3        &        333.7\\
11--12    &   103.2      &           102.0        &        111.3\\
12--13   &     29.42      &           31.93       &         31.05\\
13--14  &       9.210     &            9.480      &           ---\\
\noalign{\vspace{0.1cm}}
\hline
\end{tabular}
\end{center}
\footnotesize{The time intervals needed for the innermost shell to increment
its density of one order of magnitude, in the NODM, fiducial and maximal run. The highest density
is absent in the M1 case because the run has been stopped at smaller density, see text.}
\end{table}

\subsubsection{The maximal run}

%%%%%%%%%%%%% FIGURE 8 %%%%%%%%%%%%%
\begin{figure}
\centering
\includegraphics[angle=0,width=0.48\textwidth]{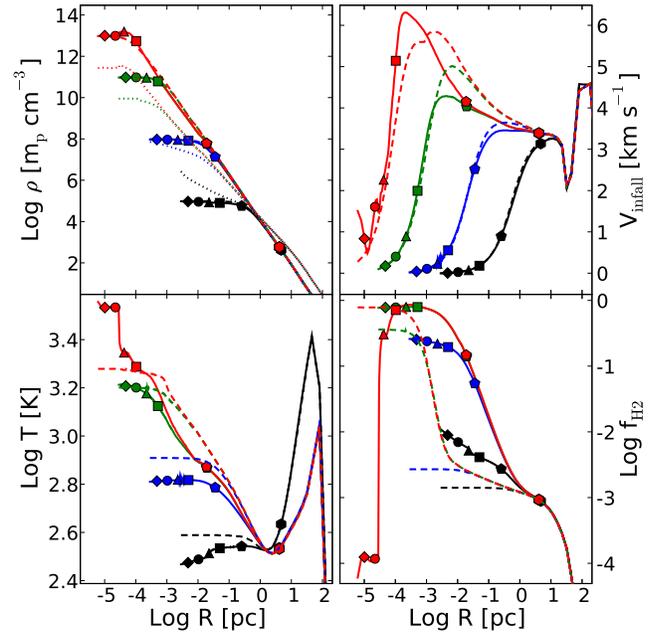}
\caption{Same as in Fig.~\ref{fig:profilo_r} but 
comparing runs M1 (maximal) and NODM (control). Solid lines refer to
gas quantities in the M1 run, dashed lines to gas quantities in the
NODM control run. All the symbols and lines are the same as in
Fig.~\ref{fig:profilo_r}, apart from the fact that the red lines refer
to a central gas density $n_{\rm c}$=10$^{13}\,\percmc$ (rather than
10$^{14}\,\percmc$); see text for comments.}
\label{fig:profilo_rmax}
\end{figure}
%%%%%%%%%% FIGURE 8 - END %%%%%%%%%%

Figure~\ref{fig:profilo_rmax}, analogue to
Fig.~\ref{fig:profilo_r}, compares the control run to
the maximal run, rather than to the fiducial run; 
in this case the highest central density we plot is $n_{\rm
c}=10^{13}\,\percmc$ (rather than $10^{14}\,\percmc$), because the
maximal run was stopped before reaching $n_{\rm c}=10^{14}\,\percmc$.

The top-left panel shows that even in this case DMAs do not alter the
self-similar part of the collapse; however, it should be remarked that
the highest density stage is starting to deviate from the expected
self-similar profile: in fact, the density profile within the core is
not flat, but there is a maximum at the edge of the core.

This is more clear when we examine the other panels: the lines
referring to the three stages with $n_{\rm c}\le 10^{11}\,\percmc$ are
qualitatively similar to their counterparts in
Fig.~\ref{fig:profilo_r}; but the lines representing the highest
density stage are clearly different, and we will focus on them.

In the lower panels we can see a huge drop in the \HH abundance near
the centre of the protostar, and a corresponding discontinuity in the
gas temperature; both could be expected by our discussion in the
previous sections, even if here we get a further piece of information
about the size of the region where the \HH was dissociated ($R\lesssim
3\times10^{-5}\,{\rm pc}$, enclosed mass $\sim
3\times10^{-2}\,\msun$). Also the velocity panel is affected: this is
the only case where the infall velocities of the run with DMAs
significantly exceed those in the control run in an extended region;
furthermore, we can see that there are secondary peaks in velocity at
low radii (in particular one is coincident with the edge of the region
where \HH was dissociated), which are likely to result from the
propagation of the density wave which was generated by the
re-expansion of the core which occurred when $n_{\rm
c}\simeq10^{12}\,\percmc$.

All these properties indicate that the core of this object is
approaching hydrostatical equilibrium, even if it is not quite there.
It is interesting to notice that the innermost
regions, of enclosed mass between 1M$_\odot$ and 10M$_\odot$, are
infalling with higher velocity than in the NODM case, whereas between
10M$_\odot$ and 10$^3$M$_\odot$ the collapse is slowed, when central
density is $n_{\rm c}$=10$^{12}\,\percmc$. Again, this is equivalent
to what we have observed for the fiducial run, and shown in Table
\ref{tab:collapsetime}, and the feature only becomes more pronounced
in the case of the maximal run.

%######
\section{Discussion}
\label{sec:discus}
%######
We have studied the effects of self-annihilating Dark Matter on the collapse of the gas structures harboring the formation of the first stars, known as Pop III.
For the first time in the literature we follow self-consistently the evolution of the Dark Matter profile as a  consequence of the gravitational drag 
of the collapsing gas and include the feedback of  energy injection by DM annihilations on the chemistry of the gas,
in the yet unexplored regime between the virialization of the halo and the formation of a hydrostatic core. We have explored the effects of DMAs by 
spanning a range of masses and annihilation cross sections around the values of the Vanilla WIMP scenario, namely those able to reproduce the relic 
abundance of DM with a thermal decoupling, finding similar results but with variations in the details and onset times of the different 
phases that we are to summarize.

In the following, quoted numbers refer to our fiducial case ($m_{\rm
DM} c^2=100\,{\rm GeV}$, $\sigmav = 3 \times 10^{-26}\,{\rm
cm^3\,s^{-1}}$, $\kappa = 0.01\,{\rm cm^2\,g^{-1}}$).  As expected,
heavier WIMPs (or smaller self-annihilation rates, or lower opacity
since the energy injection term is proportional to $\sigmav$/m$_{\rm DM}$
and depends also on $\kappa$) effects become relevant at later times
during the collapse, the same DMA rate being achieved at higher DM
densities.  Lighter particles (or higher $\sigmav$, or higher
$\kappa$) produce the same effects at earlier times (and therefore
smaller gas densities). This general scenario is robust with respect
to parameter variations within physically acceptable ranges. A few key
points of our study are worth some emphasis: {\it (i)} As known from
existing literature (RMF07), before the virialization of the halo, at
gas density $n_{\rm c} \lesssim 10^3\,\percmc$ (and therefore
extremely small gas opacity) DMAs do not sensibly affect any gas
process; {\it (ii)} between halo virialization and a gas density
$n_{\rm c} \sim 10^{11}\,\percmc$ DMAs contribute mainly through
indirect feedback effects: the free electron floor created by the
ionizations induced by DMAs catalyzes \HH formation. In turn,
molecular \HH provides more cooling to the cloud than in the standard
case (without DMAs) and the temperature of the cloud {\it decreases}
as a consequence of DMAs; {\it (iii)} finally, at $n_{\rm c} \gtrsim
10^{11}\,\percmc$ the DMA heating rate becomes equal to the gas
cooling rate. To a first approximation this leads to a balance between
losses and gain, which we dubbed as the ``critical point''.

Our results generally agree with previous ones: as the semi-analytical estimates of SFG08 and the analytical study (based on simulation data) from 
Natarajan, Tan and `O Shea (2009), we confirm the existence of a ``critical point''. We also find that the equality of \HH cooling vs DMA heating terms 
takes place (in the innermost shell) at central densities of approximately $n_{\rm c} \sim 10^{12}\,\percmc$, details depending on 
parameters (DM mass, opacity coefficient etc. see previous section, and following discussion), in agreement with the above mentioned studies.
This is particularly relevant as our analysis is the first fully numerical, self-consistent study including a reasonably accurate treatment of 
radiative transfer allowing to reach the critical point and beyond it in at least three cases.

It had been previously suggested 
%on the basis of empyrical speculations 
(SFG08) that, after reaching of the ``critical point'', the collapse
would stop and the whole structure stall, thus generating a new type
of celestial object.  With our numerical simulations we have accessed
this stage of the collapse, finding that the system does {\it not}
halt its collapse in three cases out of four; in the only case where
the collapse stops, this happens far after the critical point was
reached, and its duration is very short ($\sim 3\,{\rm yr}$) after
which the gas restarts contracting.  Moreover, the DM
parameters for which the astrophysical system finds important changes
after the critical point are actually strongly disfavored by DM
constraints based on local and primordial Universe observations.
By changing the DM mass or self-annihilation rate, 
the scenario we have described does not change
qualitatively, within the range of values studied and the 
physical regime we have accessed with our simulations.

While stressing again that when our objects 
reach the critical point
the structure does {\it not} halt the collapse, 
and instead it continues its 
evolution toward the formation of a protostellar core,
it is also useful to comment on the alteration of
the dynamics of the collapse induced by DMAs.

The duration of the gas collapse from halo virialization down to
densities of $n_{\rm c} \sim10^{14}\,\percmc$, in presence of DMAs appears
to be {\it shorter} (especially in the first phases) than in the
standard case (by about 1 per cent) of the total collapse time. 
However we point out that the change
is a small fraction of the total, and that the effects of DMAs on
collapse time are very difficult to quantify. In fact, the DMAs effect
can indeed accelerate the collapse in some shells, and decelerate it in
others, thus making it difficult to obtain a homogeneous definition of
time delay. For example, the reduction of the infall velocity of shells
enclosing $\sim 10^2\,\msun$ implies that the collapse time of these
shells is longer (by 10--20 per cent) than in the control (NODM) case.

We feel confident in stressing, however, that the shells which are
mostly affected by the infall time decrease ($\sim$20\% of the infall
velocity) are placed between 10$^2$M$_\odot$ and 10$^3$M$_\odot$, and
would become part of a hydrostatic protostar only in the final stages
of the collapse. It is not clear whether these shells will eventually end
up in the hydrostatic core (Omukai \& Palla 2003), however it is
reasonable to expect that such modification will not alter the {\it
total} time for the formation of the star by more than $\sim$20 per
cent.
%####
\subsection{Hints about the hydrostatic core}
%####

Even if our runs could not reach the regime when a hydrostatic core
forms (with the partial exception of the maximal run, where we
probably got close), it is worth to examine the likely consequences
of our results for the further evolution of the protostar in presence
of DMAs.

We start reminding that in the standard case (without DMAs) the
simulations of R02 show
that when the protostar becomes opaque to continuum radiation (at
$n_{\rm c}\simeq 10^{16}\,\percmc$), the formation of the hydrostatic
core is delayed by the thermostatic effects of \HH dissociations, so
that the core actually forms when $n_{\rm c}\gtrsim 10^{20}\,\percmc$.

However, at the end of our runs the central \HH abundance in the
maximal and sub-maximal runs is $\le 0.1$; even in the
fiducial run, the \HH abundance, while still high, is decreasing much
earlier than in the standard case; this is very different from the
results of R02, where $f_{\rm H2}\sim 1$ up to $n_{c}\gtrsim
10^{16}\,\percmc$. Because of the lower amount of H$_2$, it is
reasonable to expect that in the cases with DMAs the delaying effects
of \HH dissociation are absent, or smaller than in the standard
case. Then, the hydrostatic core would form at lower densities (e.g.,
$n_{\rm c}\sim10^{16}\,\percmc$, if the protostar becomes optically
thick to continuum radiation at the same density as in the R02 runs),
and its initial mass would be larger (probably in the 0.01--0.1
$\msun$ range, rather than $\simeq 0.003\,\msun$).

However, we point out that R02 found that in the case with no DMAs the
mass of the hydrostatic core grows very fast, reaching $0.1\,\msun$ in
less than 3 years (see e.g. their Fig. 6): then, the difference in the
initial size is relatively unimportant. It is probably more relevant
to note that the lower temperatures and infall velocities of the
layers outside the hydrostatic core imply that in the cases with DMAs
the accretion rate might be slower than what was found by R02.

\section{Conclusions}
We have studied the effects of WIMP Dark Matter Annihilations (DMAs) on the evolution of primordial gas clouds 
hosting the first stars. We have followed the collapse of gas and DM within a $10^6{\rm M}_{\odot}$ halo virializing 
at redshift $z = 20$, from $z=1000$ to slightly before the formation of a hydrostatic core, properly
including gas heating/cooling and chemistry processes induced by DMAs, and exploring the dependency of
the results on different parameters (DM particle mass, self-annihilation cross section, gas opacity, feedback strength). Independently of such parameters, when the central baryon density, $n_{\rm c}$, is lower than the critical density, $n_{\rm crit} \approx 10^{9-13}\,\percmc$, corresponding to a model-dependent balance between DMA energy input and gas cooling rate, DMA ionizations catalyze an increase in the \HH abundance by a factor $\sim 100$. The increased cooling moderately reduces the temperature (by $\approx 30$ per cent) but does not significantly reduce the fragmentation mass scale. For $n_{\rm c} \ge n_{\rm crit}$, the DMA energy injection exceeds the cooling, with the excess heat mainly going into \HH dissociations.  In the presence of DMA the transition to the continuum dominated cooling regime occurs earlier and generally is not associated with abrupt temperature variations. In conclusion, no significant differences are found with respect to the case without DMAs; in particular, and contrary to previous claims, the collapse does not stall and the cloud keeps contracting even when $n_{\rm c}\gg n_{\rm crit}$. Our simulations stop at central densities $\approx 10^{14}\,\percmc$, and cannot follow the hydrostatic core formation, nor its accretion. At the final simulation stage, the lower temperature/infall velocity of the layers enclosing a mass of $\approx 10^2 M_{\odot}$ suggest that DMAs might lead to slightly longer stellar formation timescales, with a possible $\approx 20$ per cent increase over models without DMAs.
The latter finding strengthens our conclusions, although the final answer
will come from numerical simulations (hopefully also three-dimensional) 
able to address this very same problem in the yet unexplored density regime
$10^{14}\,\percmc \lesssim n_{\rm c} \lesssim 10^{18}\,\percmc$.

\section*{Acknowledgments}
F.~I. acknowledges support from MIUR through grant PRIN-2006 during the early 
stages of this work, and from the European Community research program FP7/2007/2013 
within the framework of convention \#235878. A.~B. and P.~M. acknowledge financial 
contribution from contract ASI I/016/07/0. We thank M.~Mapelli for stimulating discussions.

%\onecolumn
\appendix
\section{Applicability of the Adiabatic Contraction formalism}
One of the hypothesis underlying the AC formalism (as derived by
Blumenthal et~al. 1986) is that the orbital time of DM particles
should be short when compared to the collapse time of the baryons.  In
the case we are studying, this is a problem: in fact, the baryonic
collapse happens essentially on a free-fall time-scale, and the DM
orbital times cannot be shorter than that.

The case of a ``fast collapse'' (where the baryon collapse time is
similar to the free-fall time) was studied by Steigman et al. (1978;
see their sec. IIIb), in the assumptions that i) the baryons stand
still at $t=0$ (so that they will collapse in free-fall), ii) the DM
contribution to the total mass $M$ is small, and iii) DM particles
have a dispersion velocity $\Delta v^2_{\rm DM}$. The case can be
treated analytically, and Steigman et al. (1978) obtain that if the
radius of the sphere enclosing all the baryons goes from $R_i$ (at
time $t=0$) to $R_c$ (at time $t=t_c$), the ratio of the number
$N_{{\rm DM},c}$ of DM particles within $R_c$ (at $t_c$) to the number
$N_{{\rm DM},i}$ of DM particles within $R_i$ (at $t=0$) is roughly
proportional to $(R_c/R_i)^{3/2}$ (cfr. eqs.~31 and~32 of Steigman et
al. 1978\footnote{Note that he power law index in eq.~32 of Steigman
  et al. (1978) should be $-3/2$ rather than $-1/2$, since this is the
  value that can be inferred from the more general eq. 31.}), at least
in the limit of $R_c\ll R_i$. Then, the density $\rho_{\rm DM}$ of DM
particles should grow asymptotically as $(R_c/R_i)^{-3/2}$, or as
$\rho_{\rm b}^{1/2}$.

We can compare this behaviour to the dependence of the central DM
density ($\rho_{{\rm DM},c}$) upon the central baryonic density
($\rho_{{\rm b},c}$) which can be obtained from the AC formalism:
SPG08 find that $\rho_{{\rm DM},c} \propto \rho_{{\rm b},c}^{0.81}$,
and our calculations essentially confirm their result.  Then, it is
clear that the ``fast collapse'' slope is significantly flatter than
what can be inferred from the application ot the AC
formalism. However, there are some extra facts that need to be kept
into account.
\begin{itemize}
\item{}{For central baryonic number densities
  $n_c\lesssim10^5\,\percmc$ the DM density is larger than the
  baryonic one, and the application of the AC does not modify the DM
  density profile;}
\item{}{The ``fast collapse'' slope was obtained with the assumption
  of an uniform density profile of both the baryons and the DM,
  whereas the NFW profile of our initial DM profile is strongly
  peaked: then, the calculations of $N_{{\rm DM},c}/N_{{\rm DM},i}$ by
  Steigman et al. (1978) likely underestimate the value appropriate in
  our case; as a consequence, it is likely that the slope of the
  $\rho_{\rm DM}-\rho_{\rm b}$ relation is higher than 0.5;}
\item{}{The DM profile seen in the simulations of Abel et
  al. 2002 appears to be more consistent with the AC predictions than
  with those of the ``fast collapse'' model: in fact, the innermost DM
  density shown in their Fig.~2 is a factor $\sim 10$ lower than the
  baryonic density $n\sim10^9\,\percmc$: this is somewhat consistent
  with the difference by a factor $[n/(10^5\,\percmc)]^{1-0.81}\sim 6$
  predicted by AC, but quite smaller than the difference by a factor
  $[n/(10^5\,\percmc)]^{1-0.81}\sim 100$ predicted using the ``fast
  collapse''.}
\end{itemize}

Then, it appears that the AC formalism can be safely used at least for
densities up to $n_c\sim10^9\,\percmc$. It is quite possible (and even
likely) that at higher density it will significantly overestimate the
DM density. However, such overestimate will be large (a maximum
$\lesssim 30$ for $n_c=10^{14}\,\percmc$) but still reasonable in the
range of densities where we perform our simulations. Furthermore, we
remark that the AC approach allows a better comparison with previous
studies, and, given our conclusions, is the {\it conservative} choice.

%% However, we can still apply the results of Steigman et al. (1978), who
%% studied also the case of a ``fast collapse'' where the baryon collapse
%% time is similar to the free-fall time (see their sec. IIIb). In
%% particular, they study the collapse of an uniform, mixed sphere of
%% baryons and DM, in the assumptions that i) the baryons stand still at
%% $t=0$ (so that they will collapse in free-fall), ii) the DM
%% contribution to the total mass $M$ is small, and iii) DM particles have a
%% dispersion velocity $\Delta v^2_{\rm DM}$. The case can be treated
%% analytically, and Steigman et al. (1978) obtain that if the radius of
%% the sphere enclosing all the baryons goes from $R_i$ (at time $t=0$)
%% to $R_c$ (at time $t=t_c$), the ratio of the number $N_{{\rm DM},c}$ of DM
%% particles within $R_c$ (at $t_c$) to the number $N_{{\rm DM},i}$ of
%% DM particles within $R_i$ (at $t=0$) is (cfr. their eq.~30):
%% \begin{equation}
%% {N_{{\rm DM},c}\over N_{{\rm DM},i}} =
%% {\rm erf}\left({{2\gamma}\over z}\right) +
%% {1\over{4\sqrt{\pi}}}
%% \left\{{\left({z\over\gamma}\right)^3 - 6\left({z\over\gamma}\right) -
%%   \left[{\left({z\over\gamma}\right)^3 - 2\left({z\over\gamma}\right)}\right]
%% \exp{\left({{-4\gamma^2}\over z^2}\right) }}\right\},
%% \end{equation}
%% where $\gamma^2=[G M /(2 R_i \Delta v^2_{\rm DM})]$, and
%% $z=[2(1-R_c/R_i)/(R_c/R_i)]^{1/2}$

\label{lastpage}
\end{document}